%% file: main.tex
\documentclass[12pt]{article}

\usepackage[utf8]{inputenc}
\usepackage[T1]{fontenc}
\usepackage[english]{babel}
\usepackage{lmodern}

\usepackage[top=2.1cm,bottom=2.9cm,left=3cm,right=2.5cm,headheight=15pt]{geometry}
\usepackage{setspace}
\usepackage{indentfirst}

\usepackage{graphicx}             
\usepackage{booktabs}
\usepackage{array,multirow,makecell}
\setcellgapes{1pt}\makegapedcells
\newcolumntype{R}[1]{>{\raggedleft\arraybackslash}b{#1}}
\newcolumntype{L}[1]{>{\raggedright\arraybackslash}b{#1}}
\newcolumntype{C}[1]{>{\centering\arraybackslash}b{#1}}
\usepackage{tabularx}
\usepackage{caption}
\usepackage[rightcaption]{sidecap}
\usepackage{subcaption}
\usepackage{float}
\usepackage{wrapfig}
\usepackage{amssymb}

\usepackage{xcolor}
\usepackage{ulem}

\usepackage{siunitx}
\usepackage{physics}

\usepackage{listings}
\usepackage{abstract}
\usepackage{titling}

\usepackage[backend=biber,style=numeric,sorting=none]{biblatex}
\usepackage{xurl}
\AtBeginBibliography{\sloppy}

\usepackage{hyperref}

\begin{document}





\input{part1}

\input{part3}

\input{part4}

\input{part5}


\printbibliography[heading=bibintoc,title={References}]

\end{document}

%% file: part1.tex
\title{\Large\textbf{Frequency Chirping of Energetic-Particle-Driven\\
Geodesic Acoustic Modes in Tokamaks}}

\author{R. Wu$^{1}$, A. Biancalani$^{1}$, M. V. Falessi$^{2}$, D. Gossard$^{1}$, R. Ivanov$^{3,1}$,\\ P. Lauber$^{4}$, X. Wang$^{4}$, and F. Zonca$^{2,5}$\\
\small ${}^1$ De Vinci Higher Education, De Vinci Research Center, Paris, France\\
\small ${}^2$ Center for Nonlinear Plasma Science and C.R. ENEA Frascati, Via E. Fermi 45,\\
\small 00044 Frascati, Italy\\
\small ${}^3$ Laboratoire de Physique des Plasmas, CNRS, Université Paris Saclay, Ecole Polytechnique,\\
\small Sorbonne Université, Observatoire de Paris, F-91120\\
\small ${}^4$ Max Planck Institute for Plasma Physics, 85748 Garching, Germany\\
\small ${}^5$ Institute for Fusion Theory and Simulation and Department of Physics,\\
\small Zhejiang University, Hangzhou 310027, China\\ \\
}


\maketitle

\begin{abstract}
A suprathermal population of ions is present in tokamak plasmas due to external heating mechanisms and fusion reactions. These energetic particles (EPs) can drive waves unstable, via inverse Landau damping.
An example is the energetic-particle-induced geodesic acoustic mode (EGAMs). In this work, investigate the nonlinear dynamics of EGAMs by means of global gyrokinetic simulations with the particle-in-cell code ORB5. In particular, we study the nonlinear evolution of the frequency, known as "frequency chirping".


To investigate the underlying phase-space dynamics, Phase Space Zonal Structure (PSZS) diagnostics are employed. By identifying the resonance energy from the phase-space distribution and reconstructing the corresponding mode frequency, an independent estimate of the frequency evolution is obtained. The reconstructed frequencies show good agreement with those extracted directly from continuous wavelet transform analysis of the electric-field signal, establishing a direct correspondence between observable frequency chirping and the nonlinear evolution of resonant energetic particles in phase space.




\end{abstract}

\section{Introduction}

Zonal (i.e. axisymmetric) flows (and their associated zonal fields, ZFs), are mesoscale organized structures often observed in tokamaks in the presence of turbulence~\cite{Hasegawa79,Chen00}.
The characteristic polarization has the flow going along the poloidal direction, and a dependence on the minor radius of the tokamak, i.e. the direction of non-uniformity of the equilibrium temperature and density (no dependence on the toroidal angle due to the axisymmetry). Two kinds of ZFs exist: zero-frequency zonal flows, and an oscillatory branch known as geodesic acoustic modes
(GAMs)~\cite{winsor1968}. Zero-frequency zonal flows are mainly damped by collisions, whereas GAMs are damped by both collisions and collisionless Landau damping~\cite{landau1946}.
ZFs are key ingredients in the self-organization of turbulence in toroidal plasmas. Hence, the importance of understanding their dynamics, in order to have a comprehensive model of turbulent transport.


The presence of a population of energetic (i.e. suprathermal) particles (EPs), such as those generated by neutral beam injection or fusion-born $\alpha$ particles, qualitatively modifies this picture. Through inverse Landau damping, EP populations can destabilize GAMs, giving rise to EP-driven GAMs (EGAMs)~\cite{Fu08}. EGAMs are of particular interest mainly because of their interaction with the EPs in phase space. They could also be mediators between turbulence and EPs, and therefore influence transport and confinement~\cite{Zarzoso13}.


Several contributions have been given to the construction of a theoretical model of EGAMs~\cite{Fu08,Qiu10,Qiu11,Zarzoso13,Miki15,Sasaki16,Sasaki17,Novikau20}. Due to their nature of forced-oscillations, the dynamics of EGAMs strongly depends on the time evolution of the EP population in phase space. Not only their drive decreases when the velocity-gradient is depleted due to the nonlinear EP redistribution: also the frequency can change in time, during the nonlinear phase. This phenomenon, named frequency chirping, is a signature of the intrinsic nonlinear physics characterizing the EGAM interaction with EPs.

Frequency chirping is a known feature of many instabilities in nature. In tokamaks, it has been studied in detail for example for Alfv\'en modes~\cite{Fasoli08,Pinches04,chen2016,wang2023}, and for fishbones~\cite{Brochard25}. In space, chorus waves have been also found to present a similar dynamics~\cite{Tao20}.
For EGAMs, frequency chirping has been observed in numerical simulations (see for example Ref.~\cite{WangPRL13,biancalani2017,biancalani2018}) and compared with experiments (for example in Ref.~\cite{Novikau20}).
Although GAMs and EGAMs are oscillations which exist only due to the magnetic field curvature~\cite{winsor1968}, and therefore only in a 2D shaped magnetic equilibrium like the tokamak, nevertheless, the EGAMs drive has a strong analogy with the beam plasma instability (BPI)~\cite{Qiu10,biancalani2017}. The most basic model of a BPI can be explained as a 1D physics phenomenon. The beam plasma instability occurs when a plasma wave (also known as Langmuir wave) taps the free energy of a beam of energetic electrons, and becomes unstable due to inverse Landau damping~\cite{Oneil68}. As a consequence, not only the linear excitation, but also the nonlinear saturation level of the EGAMs scales similarly to the BPI~\cite{biancalani2017,biancalani2018}.

The dynamics of ZFs can be studied in the more general framework of renormalized nonlinear equilibria, dubbed Zonal State \cite{Falessi19, Falessi23, Qiu25}. Zonal states  consist of ZFs, i.e. the toroidally symmetric electromagnetic response, and Phase-Space zonal structures, i.e. the collisionlessly undamped, orbit-averaged particle response extracted in phase space, which evolve self-consistently with fluctuations, sources and collisions. In this sense, EGAM saturation and frequency chirping may be viewed as the evolution of the Zonal State.
The questions that did not have a comprehensive answer so far are: how does the frequency chirping of EGAMs develop during the nonlinear phase in relation to the resonant EPs? And how can this be explained in terms of PSZS evolution?
This is the main focus of the present work.

The main numerical tool used here is the particle in cell code ORB5~\cite{orb5}.
ORB5 is a global gyrokinetic code, written originally for turbulence studies. Gyrokinetics is the framework which is valid in the regime of frequencies which are much smaller than the ion cyclotron frequencies. In this regime, the fast gyro-motion of the particles can be averaged out, thus reducing the original 6D model for the particle motion in phase space to a 5D model. This allows faster numerical simulations. ORB5 is a multi-species code, and therefore it can be used to study EP-driven instabilities. We choose an EP distribution function  of the shape of a bump-on-tail, i.e. a maxwellian shifted in the direction of the parallel velocity. For simplicity, the electrons are treated adiabatically here, as we wish to focus on the EP dynamics (following the prescription of Ref.~\cite{biancalani2017}).
After validating the numerical framework with standard benchmarks, we investigate how the linear growth rate, nonlinear saturation level, and frequency chirping of EGAMs depend on the energetic particle concentration. The quadratic scaling of the saturation level with the linear growth rate is recovered, as given in Ref.~\cite{biancalani2017}. Regarding the frequency chirping rate, we obtain a linear scaling with respect to the linear growth rate.

The paper is organized as follows. 
The model equilibrium and numerical setup are described in Section~\ref{sec:model and eq}.
The nonlinear frequency chirping of EGAMs is studied in Section~\ref{sec:NL dyn chirping}, where the scalings of the EGAM frequency chirping are also shown. The interpretation of the  frequency chirping in terms of PSZS nonlinear evolution is given in Section~\ref{sec:NL dyn of PSZS}. A summary and discussion are given in Section~\ref{sec:summary}.

%% file: part3.tex
\section{Model and Equilibrium}
\label{sec:model and eq}

\subsection{GAMs and EGAMs}

GAMs are axisymmetric oscillations of the zonal
$\vb{E}\times\vb{B}$ flow, whose restoring force originates from the geodesic curvature
of the magnetic field \cite{winsor1968}. When a flux-surface-averaged radial electric field perturbation
is generated, the associated $\vb{E}\times\vb{B}$ flow induces a compression of the
plasma density on the flux surface. The resulting density perturbation drives parallel
currents, which react back on the radial electric field, forming a closed feedback loop
that sustains oscillatory dynamics.

In thermal plasmas, GAMs oscillate at an acoustic frequency of order
\begin{equation}
\omega_{\mathrm{GAM}} =
\frac{\sqrt{2}\, c_s}{R_0},
\label{eq:omega_gam}
\end{equation}
where $c_s$ is the ion sound speed and $R_0$ is the major radius of the tokamak.
Kinetic effects, in particular ion Landau damping, lead to strong attenuation of GAMs
in Maxwellian plasmas.

When energetic particles (EPs) are present, the dynamics change qualitatively. Non- Maxwellian EP velocity distributions, such as the double bump-on-tail profiles, can provide a positive velocity-space gradient near resonance, enabling energy transfer from particles to the wave through inverse Landau damping. Consequently, GAMs can become unstable, giving rise to EGAMs \cite{biancalani2017}. In nonlinear simulations, the growth is eventually arrested and the mode saturates at a finite amplitude; the saturation physics is discussed in Sec. 4.

\subsection{Tokamak Equilibrium and Plasma Profiles}

The simulations are carried out in a simplified tokamak equilibrium with circular magnetic surfaces and no Shafranov shift. The equilibrium parameters are chosen consistently with those adopted in previous studies of the dynamics of EGAM~\cite{biancalani2017,biancalani2018}. The major and minor radii are set to $R_0=\SI{1}{m}$ and $a=\SI{0.3125}{m}$, respectively, while the magnetic field strength on the axis is $B_0=\SI{1.9}{T}$. A flat safety-factor profile with $q=2$ is assumed throughout the plasma volume.
The thermal plasma is characterized by flat density and temperature profiles. A hydrogen plasma with equal ion and electron temperatures ($T_e=T_i$) is considered. The normalized ion-sound Larmor radius is set to $\rho^\ast=\rho_s/a=1/128$, where $\rho_s=c_s/\Omega_i$ is the ion-sound Larmor radius, $c_s=\sqrt{T_e/m_i}$ is the sound speed, and $\Omega_i$ is the ion cyclotron frequency. For a hydrogen plasma, this corresponds to $T_i=T_e=2060~\mathrm{eV}$, $\Omega_i=1.82\times10^8~\mathrm{rad,s^{-1}}$, and $c_s=4.44\times10^5~\mathrm{m,s^{-1}}$. Since $T_e=T_i$, the ion thermal velocity $v_{ti}=\sqrt{T_i/m_i}$ coincides with the sound speed, i.e. $v_{ti}=c_s$. The corresponding sound frequency is $\omega_s=\sqrt{2}~ c_s/R_0=6.28\times10^5~\mathrm{rad ~s^{-1}}$.

\begin{figure}[H]
\centering

\begin{subfigure}[b]{0.48\textwidth}
\centering
\includegraphics[width=\textwidth]{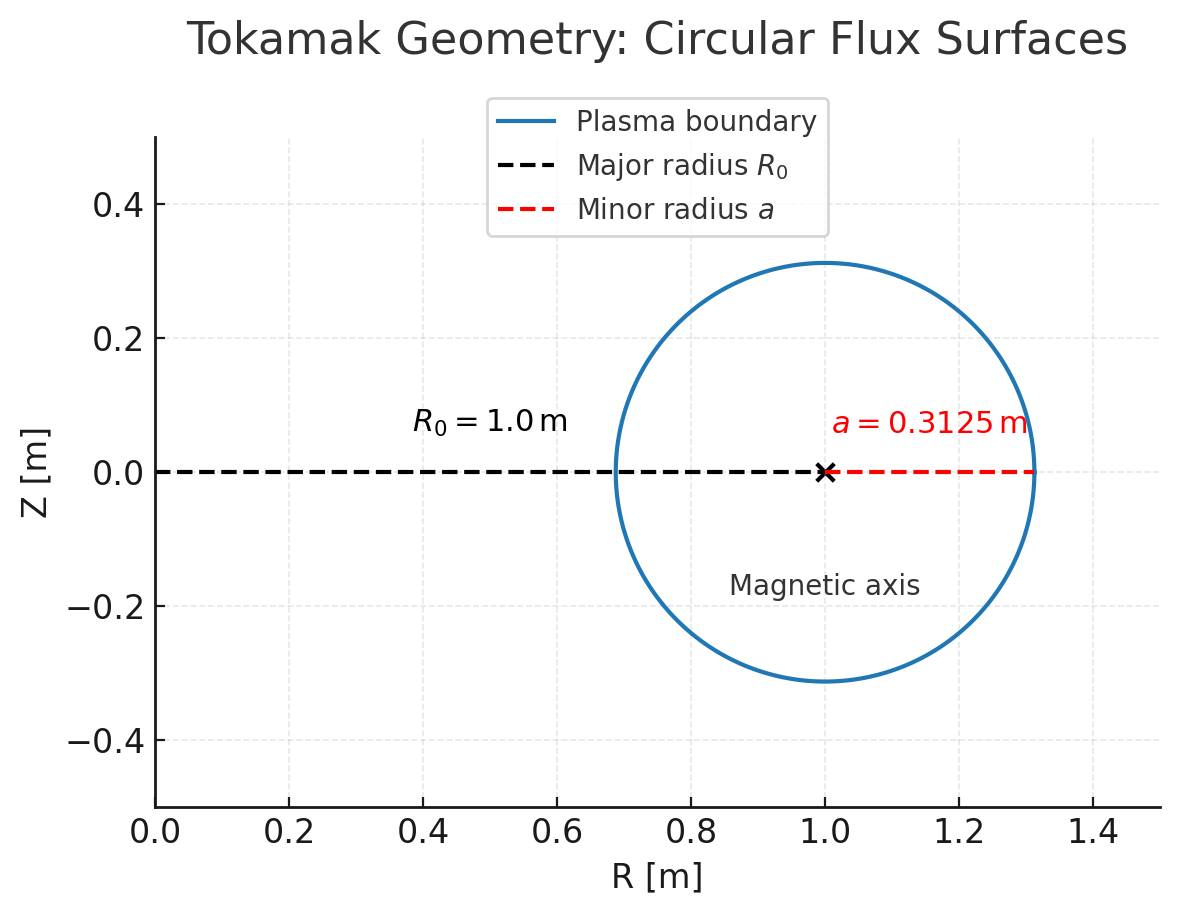}
\caption{Tokamak geometry.}
\label{fig:geometry}
\end{subfigure}
\hfill
\begin{subfigure}[b]{0.48\textwidth}
\centering
\includegraphics[width=\textwidth]{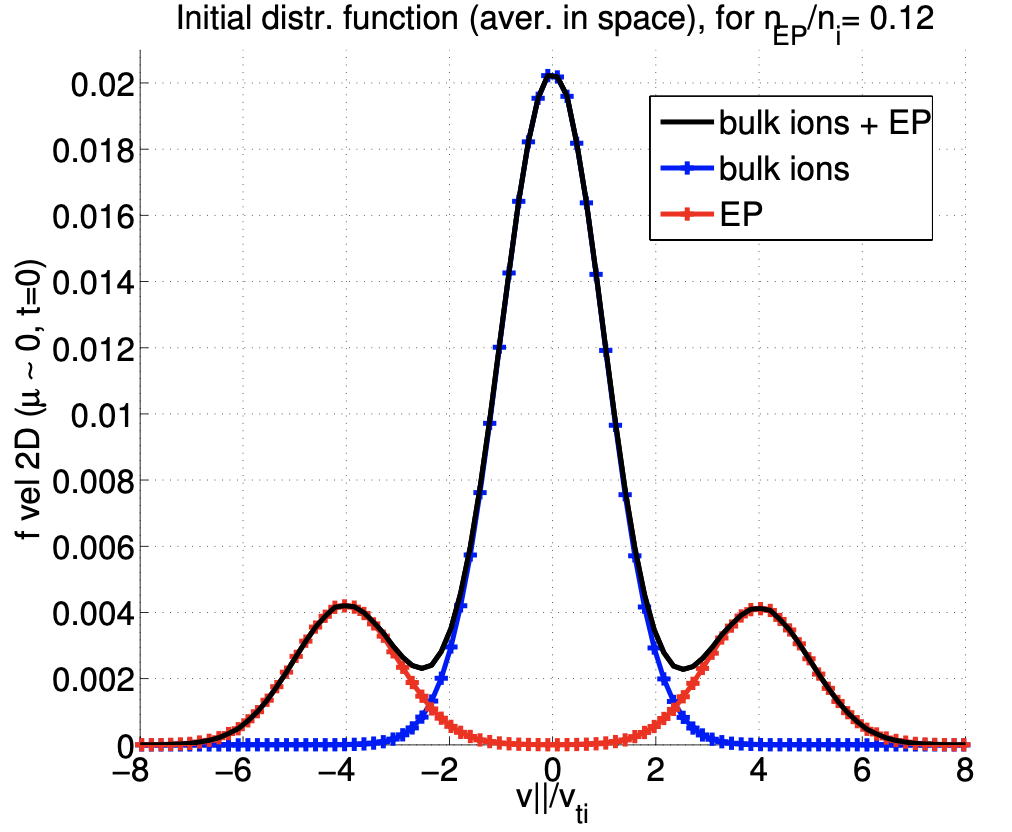}
\caption{Initial EP double bump-on-tail distribution function.}
\label{fig:initial_distribution}
\end{subfigure}

\caption{
(a) Tokamak geometry with $R_0=1~\mathrm{m}$ and $a=0.3125~\mathrm{m}$.
(b) Initial EP double bump-on-tail distribution function for a simulation with $n_{EP}/n_i = 0.12$, $v_{bump}/v_{ti} = 4$. Reproduced from Biancalani et al. (2017) \cite{biancalani2017}.
}
\label{fig:model_setup}
\end{figure}

Energetic particles (EPs) are initialized using a symmetric bump-on-tail distribution
function shown in \autoref{fig:initial_distribution}, characterized by two peaks in parallel velocity at
\begin{equation}
v_\parallel = \pm v_{\mathrm{bump}}, \qquad
v_{\mathrm{bump}} = 4 v_{ti},
\end{equation}
The EP distribution depends only on the constants of motion, and the radial variations of the magnetic field are neglected in the initialization.



Neumann and Dirichlet boundary conditions are imposed on the electrostatic potential
at the inner and outer boundaries, corresponding to $s=0$ and $s=1$, respectively.
As the EP concentration is increased, the system transitions smoothly from a damped
GAM to an unstable EGAM.

The energetic-particle population is characterized by the density ratio $n_{EP}/n_i$, where $n_{EP}$ and $n_i$ denote the energetic-particle and thermal-ion densities, respectively.

\subsection{The ORB5 Gyrokinetic Code}

The simulations are performed using the global gyrokinetic particle-in-cell (PIC) code ORB5~\cite{orb5, jolliet2007}, which solves the gyrokinetic Vlasov--Maxwell equations in five-dimensional phase space.

In this work, the electrostatic and collisionless version of ORB5 is employed, with electrons treated adiabatically. Within this framework, the gyrocenter trajectories of the energetic particles are evolved self-consistently together with the electrostatic potential, providing a kinetic description of resonant wave--particle interactions.

The kinetic treatment is essential for the investigation of EGAM dynamics, since both the linear excitation through inverse Landau damping and the subsequent nonlinear evolution are determined by the redistribution of energetic particles in phase space. In particular, ORB5 allows one to follow the evolution of the energetic-particle distribution function, while dedicated PSZS diagnostics~\cite{Bottino22} are employed in the post-processing stage to investigate the underlying phase-space dynamics.

The energetic-particle markers are evolved along trajectories including the perturbation associated with the EGAM electric field, whereas the bulk-ion markers are evolved along unperturbed trajectories. As a consequence, wave--wave coupling effects are neglected and the nonlinear dynamics are entirely determined by resonant wave--particle interactions.

\subsection{Signal Processing and Diagnostic Methods}

To analyze the temporal evolution of EGAMs, the flux-surface-averaged electrostatic
potential $\bar{\phi}(t)$ is first extracted from the simulations. The  
radial electric field is then computed as
\begin{equation}
E_r(t) = -\partial_r \bar{\phi}(t),
\end{equation}
which serves as the primary diagnostic quantity throughout this work.

The linear growth rate $\gamma_L$ is determined by fitting the exponential growth
phase of the signal on a semi-logarithmic scale. A linear regression of the early-time
interval provides a direct estimate of $\gamma_L$.

To investigate the nonlinear frequency dynamics, the time–frequency evolution of Er(t) was analyzed using the continuous wavelet transform (CWT). The CWT method provides a localized spectral decomposition, allowing the clear identification of mode frequencies as they evolve in time, and is particularly suitable for capturing the chirping behavior characteristic of EGAMs. The instantaneous frequency was obtained by following the ridge of maximum wavelet amplitude, and the chirping rate was estimated by applying a linear regression to the extracted frequency trajectory.

To complement the signal-based analysis, a diagnostic framework based on Phase Space Zonal Structures (PSZS) formalism developed by Zonca, Chen and
collaborators is employed \cite{zonca2015,Falessi19,Falessi23}. Unlike the CWT method, which extracts the frequency evolution directly from the field fluctuations, the PSZS diagnostic probes the resonant energetic-particle dynamics in phase space (see Sec.~\ref{sec:NL dyn of PSZS}).

By establishing a correspondence between the chirping features extracted from the CWT analysis and the evolution of PSZS, the nonlinear dynamics can be examined from two independent perspectives. This approach enables a direct connection between the observable frequency evolution and the underlying phase-space structures responsible for wave–particle interactions.

%% file: part4.tex

\section{Nonlinear frequency chirping of EGAMs}
\label{sec:NL dyn chirping}

\subsection{Evolution of the frequency in time}

\begin{figure}[H]
\centering
\begin{subfigure}{0.48\textwidth}
    \centering
    \includegraphics[width=\textwidth]{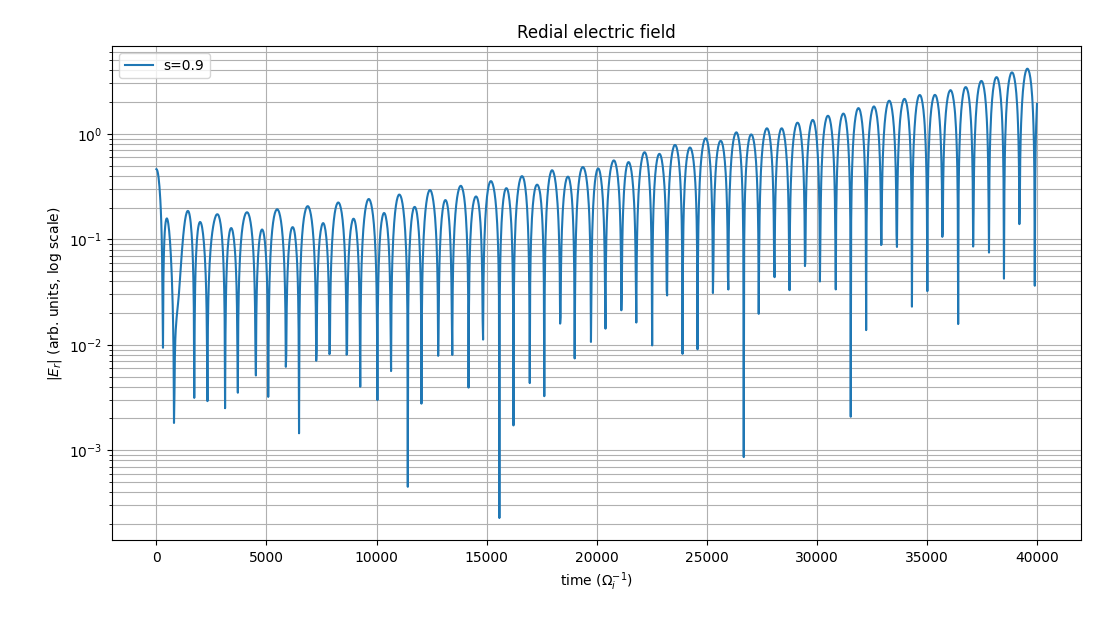}
    \caption{Linear simulation}
    \label{fig:Er_linear}
\end{subfigure}
\hfill
\begin{subfigure}{0.48\textwidth}
    \centering
    \includegraphics[width=\textwidth]{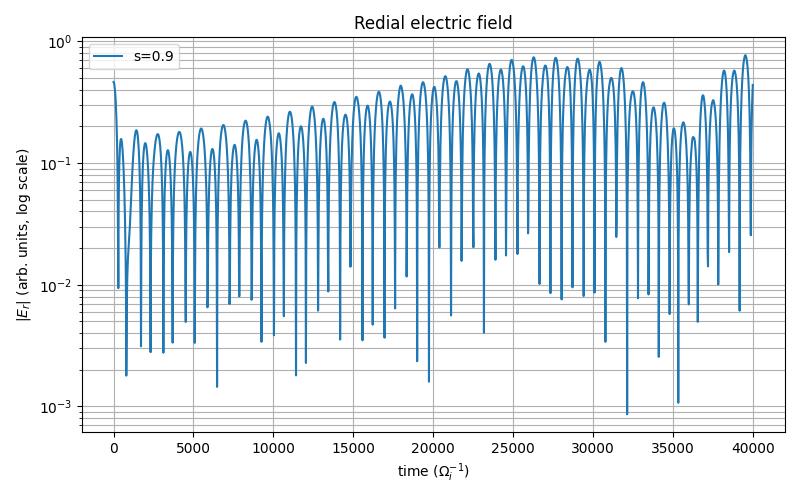}
    \caption{Nonlinear simulation}
    \label{fig:Er_nonlinear}
\end{subfigure}
\caption{Time evolution of the radial electric field $E_r(t)$ at $s=0.9$ with an EP concentration of $7\%$. 
Panel (a) corresponds to the linear simulation, while panel (b) corresponds to the nonlinear simulation.}
\label{fig:Er_lin_nonlin}
\end{figure}

To illustrate the fundamental features of EGAMs under different simulation conditions, we first show the time evolution of the radial electric field in both linear and nonlinear simulations.  

In the linear simulation (see \autoref{fig:Er_linear}), the amplitude of the radial electric field exhibits an almost perfect exponential growth. This behavior is fully consistent with linear theory, where the mode freely grows under the EP drive without being affected by nonlinear effects. Such signals are particularly suitable for the measurement of the linear growth rate $\gamma_L$, since the exponential growth region can be accurately captured by logarithmic linear fitting. It should be noted that, during this stage, the mode frequency remains constant and no chirping phenomenon is observed.  

In contrast, in the nonlinear simulation (see \autoref{fig:Er_nonlinear}), the time evolution of the radial electric field exhibits significantly different characteristics. The mode also undergoes an initial exponential growth, but once the amplitude reaches a certain level, the exponential law breaks down and the system enters the nonlinear stage. In this regime, the mode amplitude saturates at a finite level, often accompanied by frequency chirping in the electric field signal. These features are direct evidence of nonlinear wave--particle interactions.  

\begin{figure}[H]
\centering
\includegraphics[width=0.7\textwidth]{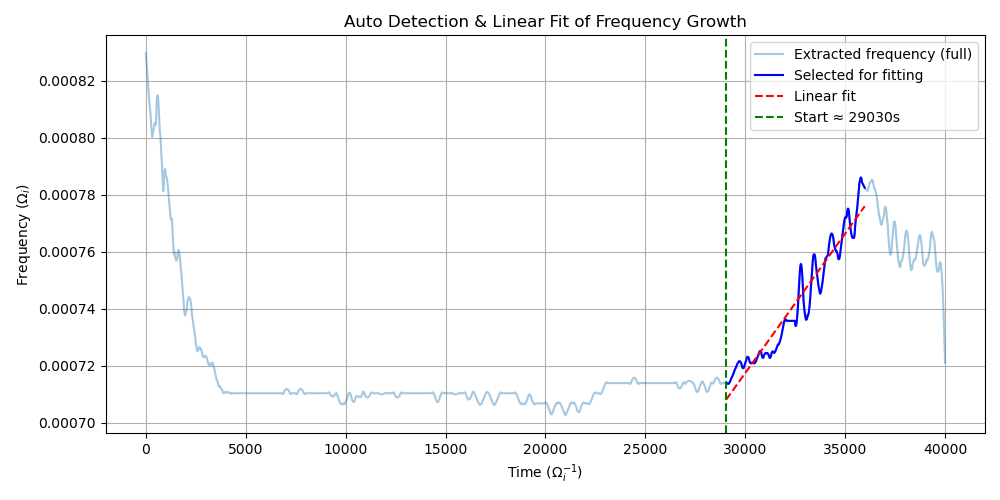}
\caption{Time--frequency representation of the $E_r(t)$ signal at $s=0.9$ with an EP concentration of $7\%$, obtained via continuous wavelet transform (CWT). One can notice that in the very beginning of the signal ($t=0$ to $t \simeq 5000 \Omega_i^{-1}$), the frequency appears to decrease. This feature is not of physical origin, but rather an artifact introduced by the CWT signal processing, and can therefore be safely neglected. 
Around $t \simeq 29000,\Omega_i^{-1}$, the frequency starts to drift, indicating the onset of chirping. 
}
\label{CWT}
\end{figure}

This conclusion is further supported by the time--frequency analysis. \autoref{CWT} presents the frequency evolution of the $E_r(t)$ signal extracted via the continuous wavelet transform (CWT). The CWT is performed using a complex Morlet wavelet (cmor1.5-1.0), with logarithmically spaced scales in the range $ \in [3, 100]$ (1000 points), providing sufficient resolution in both time and frequency domains.
The instantaneous frequency is obtained by identifying, at each time step, the ridge corresponding to the maximum wavelet amplitude. To reduce noise-induced fluctuations, the extracted frequency is smoothed using a Savitzky--Golay filter.

The apparent decrease in frequency at early times is not of physical origin but arises from boundary effects of the CWT (cone of influence). This region is therefore excluded from further analysis. 
In the initial stage of the EGAM signal ({$t \simeq 5000 \Omega_i^{-1}$}), the mode frequency remains nearly constant, corresponding to the linear growth phase of the radial electric field. However, when time approaches $t \simeq 29000\,\Omega_i^{-1}$, the frequency starts to evolve in time, indicating the onset of chirping. 



The frequency evolution during the chirping phase is not strictly monotonic, since both upward and downward excursions can occur at later times. 
Therefore, the chirping rate is evaluated only during the early nonlinear phase, namely from the onset of the frequency shift up to the saturation of the mode amplitude, where the frequency evolution is approximately monotonic (highlighted region in \autoref{CWT}). This allows for a reliable linear fit of the instantaneous frequency.


Based on these observations, in the following sections we quantitatively investigate the relations among the linear growth rate, the nonlinear saturation level, and the chirping rate, with the goal of characterizing the scaling laws governing the nonlinear dynamics of EGAMs.

\subsection{Dependence of Linear Growth Rate and Saturation Level on EP Concentration}

\begin{figure}[h]
\centering
\begin{subfigure}{0.48\textwidth}
    \centering
    \includegraphics[width=\textwidth]{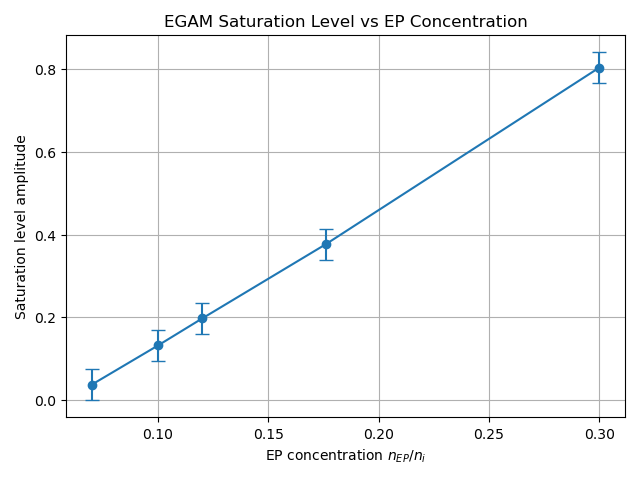}
    \caption{Saturation amplitude as a function of EP concentration.}
    \label{fig:sat_vs_nEP}
\end{subfigure}
\hfill
\begin{subfigure}{0.48\textwidth}
    \centering
    \includegraphics[width=\textwidth]{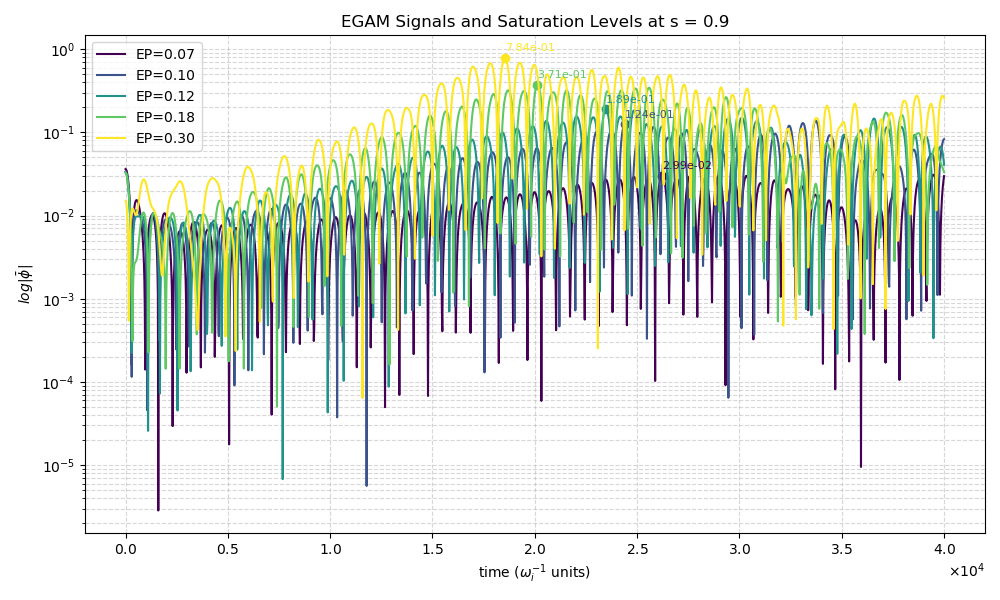}
    \caption{Time evolution of EGAM signals at different EP concentrations, with the corresponding saturation levels.}
    \label{fig:saturation_level}
\end{subfigure}
\caption{
Dependence of the EGAM saturation level on EP concentration.
(a) Saturation amplitude as a function of EP concentration $n_{EP}/n_i$.
(b) Time evolution of the radial electric field for different EP concentrations at $s=0.9$.
}
\label{fig:saturation}
\end{figure}

We first investigate the dependence of the EGAM saturation level on the EP concentration. As shown in \autoref{fig:saturation}, the saturation amplitude increases systematically with increasing EP concentration. This behavior reflects the stronger energetic-particle drive available at higher EP densities, which allows the mode to reach a larger nonlinear equilibrium amplitude before saturation.


\begin{figure}[H]
\centering
\includegraphics[width=0.5\textwidth]{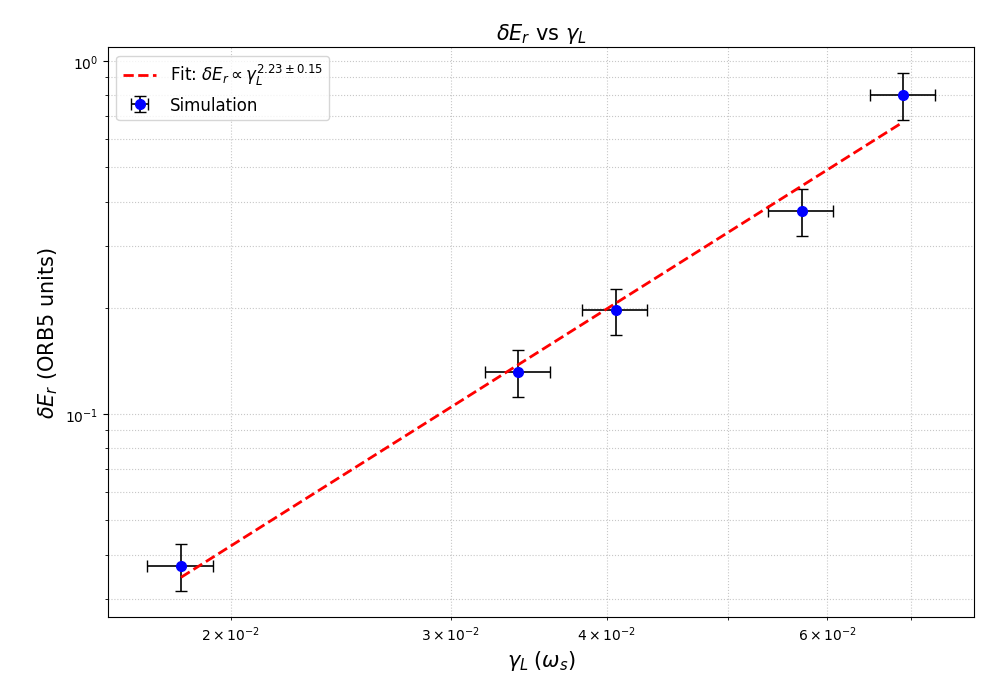}
\caption{Relationship between the linear growth rate $\gamma_L$ and the saturation level $\delta E_r$ at $ s = 0.9$ 
}
\label{fig:gamma_deltaEr}
\end{figure}



\autoref{fig:gamma_deltaEr} compares the saturation level with the linear growth rate, revealing a clear positive correlation: larger growth rates lead to higher saturation amplitudes. On logarithmic axes, the relationship is well described by a power-law fit, yielding

\begin{equation}
\delta E_r \propto \gamma_L^{2.23 \pm 0.15},
\end{equation}

which is consistent, within uncertainty, with the quadratic scaling

\begin{equation}
\delta E_r \propto \gamma_L^2
\end{equation}

measured by A.Biancalani et al.\cite{biancalani2017}.

The observed scaling is characteristic of saturation dominated by nonlinear wave--particle interactions. As the EGAM amplitude increases, resonant energetic particles become trapped and redistributed around the resonance region, progressively reducing the net energy transfer from particles to the mode. Saturation is reached when the drive and damping balance each other.


\subsection{Dependence of Chirping Rate and Saturation Level on EP Concentration}

\begin{figure}[H]
\centering
\begin{subfigure}{0.48\textwidth}
    \centering
    \includegraphics[width=\textwidth]{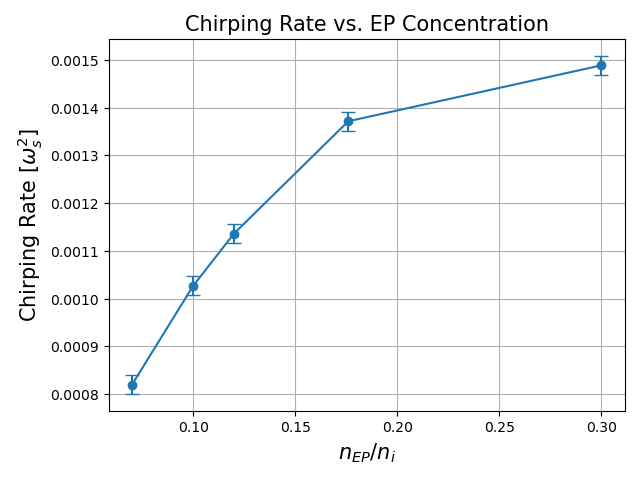}
    \caption{Chirping rate vs EP concentration $n_{EP}/n_i$ at $s = 0.9$.}
    \label{fig:chirp_vs_nEP}
\end{subfigure}
\hfill
\begin{subfigure}{0.48\textwidth}
    \centering
    \includegraphics[width=\textwidth]{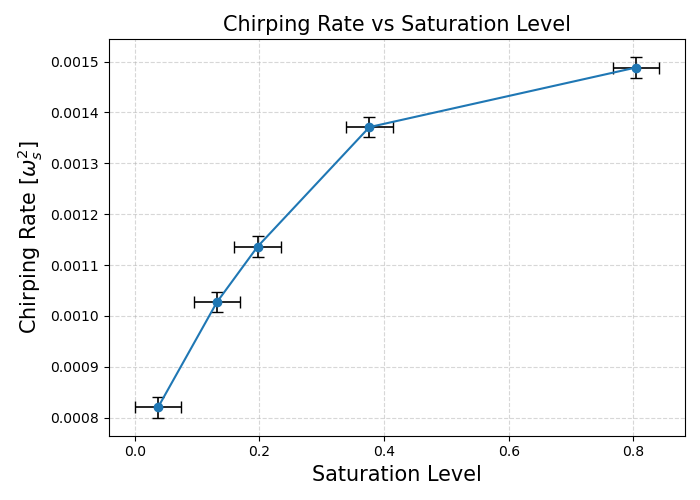}
    \caption{Chirping rate vs saturation level $\delta E_r $ at $s = 0.9$.}
    \label{fig:chirp_deltaEr}
\end{subfigure}
\caption{Dependence of chirping rate}
\label{fig:chirp}
\end{figure}

\autoref{fig:chirp_vs_nEP} shows the dependence of the chirping rate on the EP concentration. A clear systematic trend is observed: as the EP drive becomes stronger, the frequency sweeping accelerates. At low EP concentrations, the chirping rate remains relatively small in magnitude but already exhibits a pronounced sensitivity to variations in the fast-particle population. As the EP concentration is further increased, the chirping rate attains larger values, consistent with the stronger drive. 
The dependence of the chirping rate can be studied also as a function of the saturation level. The result is given in \autoref{fig:chirp_deltaEr}. Within the error bars, we can observe a trend which is approximately linear in the regime of low drive, and then a change in slope for very large drives.

The interpretation of the change of regime found in \autoref{fig:chirp_deltaEr} is given as follows. To the lowest order, we can model the EGAM behavior as determined mainly by the resonant EPs, i.e. those with velocity near 
\begin{equation}
    v_{r}= \omega_{EGAM}/qR
\end{equation}

\begin{figure}[h]
\centering
\begin{subfigure}[b]{0.48\textwidth}
\centering
\includegraphics[width=\textwidth]{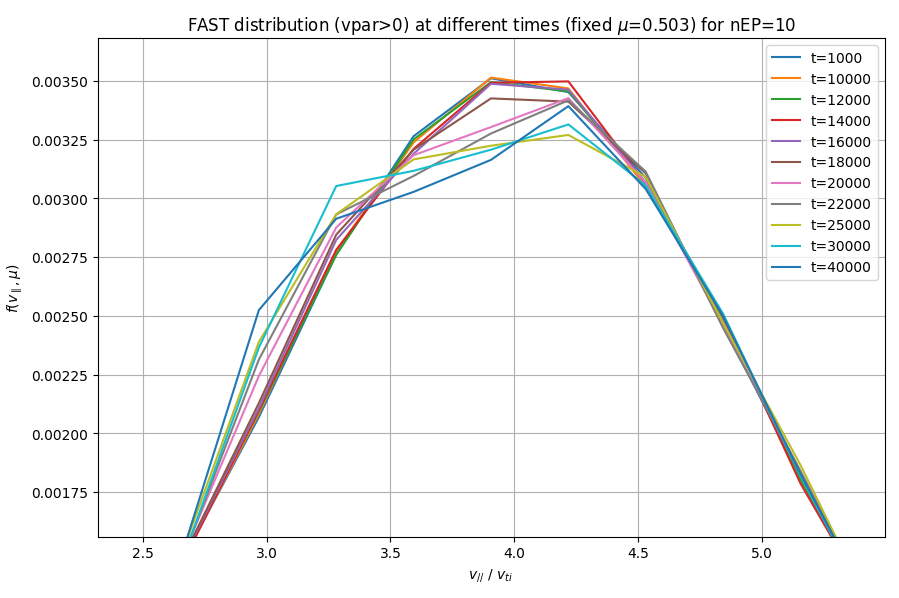}
\caption{EP distribution function}
\label{fig:distribution}
\end{subfigure}
\hfill
\begin{subfigure}[b]{0.48\textwidth}
\centering
\includegraphics[width=\textwidth]{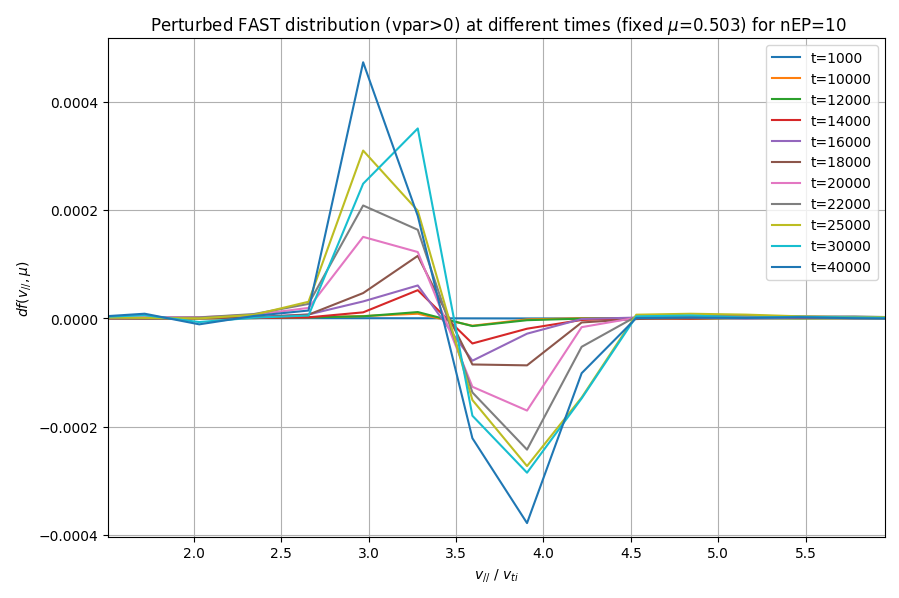}
\caption{Perturbed EP distribution function}
\label{fig:df}
\end{subfigure}
\caption{
Time evolution of the energetic-particle distribution at fixed magnetic moment $\mu \simeq 0.0$ for the case with $n_{EP}/n_i=10\%$. (a) Distribution function. (b) Perturbed distribution function.
}
\label{fig:model_setup}
\end{figure}

In this approximation, the main mechanism of nonlinear modification of the EGAM dynamics is given simply by the redistribution of those resonant EPs in velocity space, and little effect is given by the EPs which are far from the resonance. Note that the simulations presented here were studied also in Ref.~\cite{biancalani2018} for the dynamics in velocity space. The simulations with low drive were shown to have little EP redistribution in phase space, whereas those simulations with large EP drive had a total modification of the EP distribution function during the nonlinear saturation (and not only near the resonance) as shown in \autoref{fig:distribution}. In \autoref{fig:chirp_deltaEr}, the large-drive regime corresponds to EP concentrations of 17.5\% and 30\%, i.e. the last two points on the right. In summary, we identify a low-drive regime as an approximately linear scaling of the chirping rate vs saturation level, whereas the scaling changes when we go to large-drive regime.

\begin{figure}[H]
\centering
\includegraphics[width=0.45\textwidth]{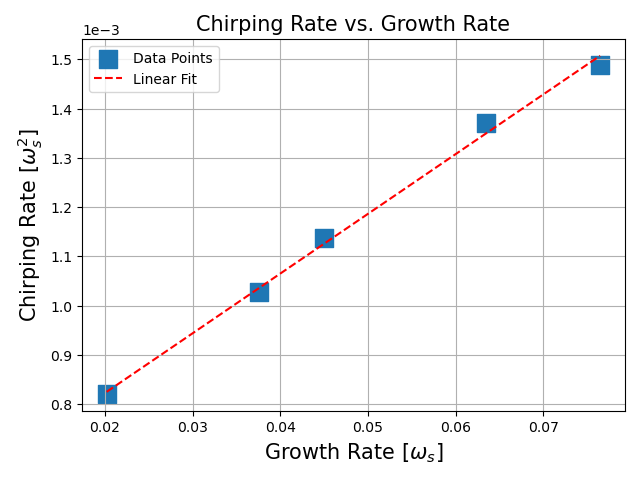}
\caption{Chirping rate versus growth rate.}
\label{growth_vs_chirp}
\end{figure}

For completeness, we also study the scaling of chirping rate vs linear growth rate. This is shown in \autoref{growth_vs_chirp}. A linear scaling is observed for the whole range of EP concentration considered here.


\section{Nonlinear dynamics of phase space zonal structures}
\label{sec:NL dyn of PSZS}

In addition to the frequency analysis based on the Continuous Wavelet Transform (CWT), we also employ the Phase Space Zonal Structures (PSZS) diagnostics, which was recently implemented in the global GK PIC code ORB5~\cite{Bottino22}. The PSZS diagnostic reconstructs the orbit-averaged energetic particle distribution function through a finite-element projection in phase space, allowing one to follow the slow nonlinear evolution of energetic particle redistribution driven by Alfvénic activity. Since the PSZS contains direct information on the nonlinear wave--particle interaction dynamics in phase space, it provides a complementary and physically insightful diagnostic compared with conventional field-based analyses, and is particularly useful for investigating nonlinear energetic particle transport processes.

In this section, we extract the resonance energy and its temporal evolution from the PSZS diagnostic output. The corresponding mode frequency evolution is then inferred and quantitatively compared with the frequencies obtained from the CWT analysis, in order to assess the consistency between the two independent approaches.

\begin{figure}[h]
\centering
\includegraphics[width=0.6\textwidth]{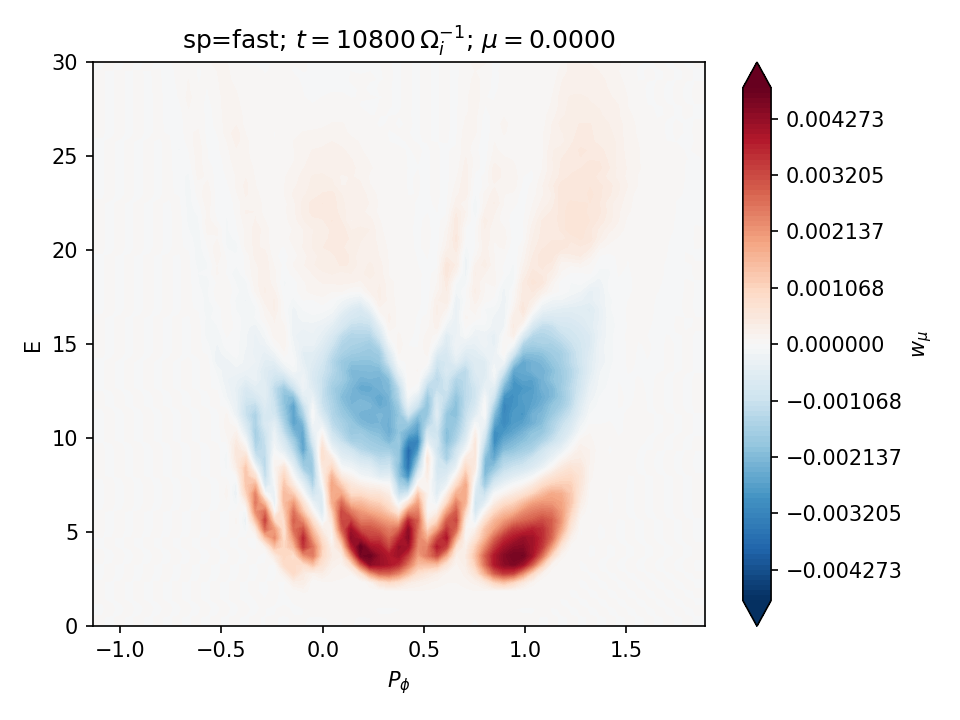}
\caption{2D phase-space projection of the perturbed distribution function $w_\mu$ for fixed $\mu = 0$ in the $(P_\phi,E)$ space obtained from PSZS diagnostics for the case with EP concentration $n_{EP}/n_i = 17.6\%$. The resonance layer is characterized by alternating positive and negative perturbations separated by a narrow near-zero layer, corresponding to the wave–particle resonance region and the associated energetic particle redistribution.}
\label{frame}
\end{figure}

The phase-space structures shown in \autoref{frame} exhibit a redistribution predominantly localized along the energy coordinate, while remaining relatively confined in the $P_\phi$ direction. Since the canonical toroidal momentum $P_\phi$ is closely related to the magnetic flux coordinate and therefore to the radial position of particle orbits, this observation indicates that the nonlinear EGAM dynamics is dominated by velocity-space redistribution rather than radial transport. 

This observation is consistent with the characteristic physics of the EGAM wave-particle interaction, where the nonlinear interaction mainly develops along the parallel velocity direction as discussed by Biancalani et al.~\cite{biancalani2018} Therefore, the alternating positive and negative perturbations separated by a narrow near-zero layer correspond to local resonant flattening of the energetic-particle distribution function in velocity space.

\begin{figure}[H]
\centering
\includegraphics[width=0.6\textwidth]{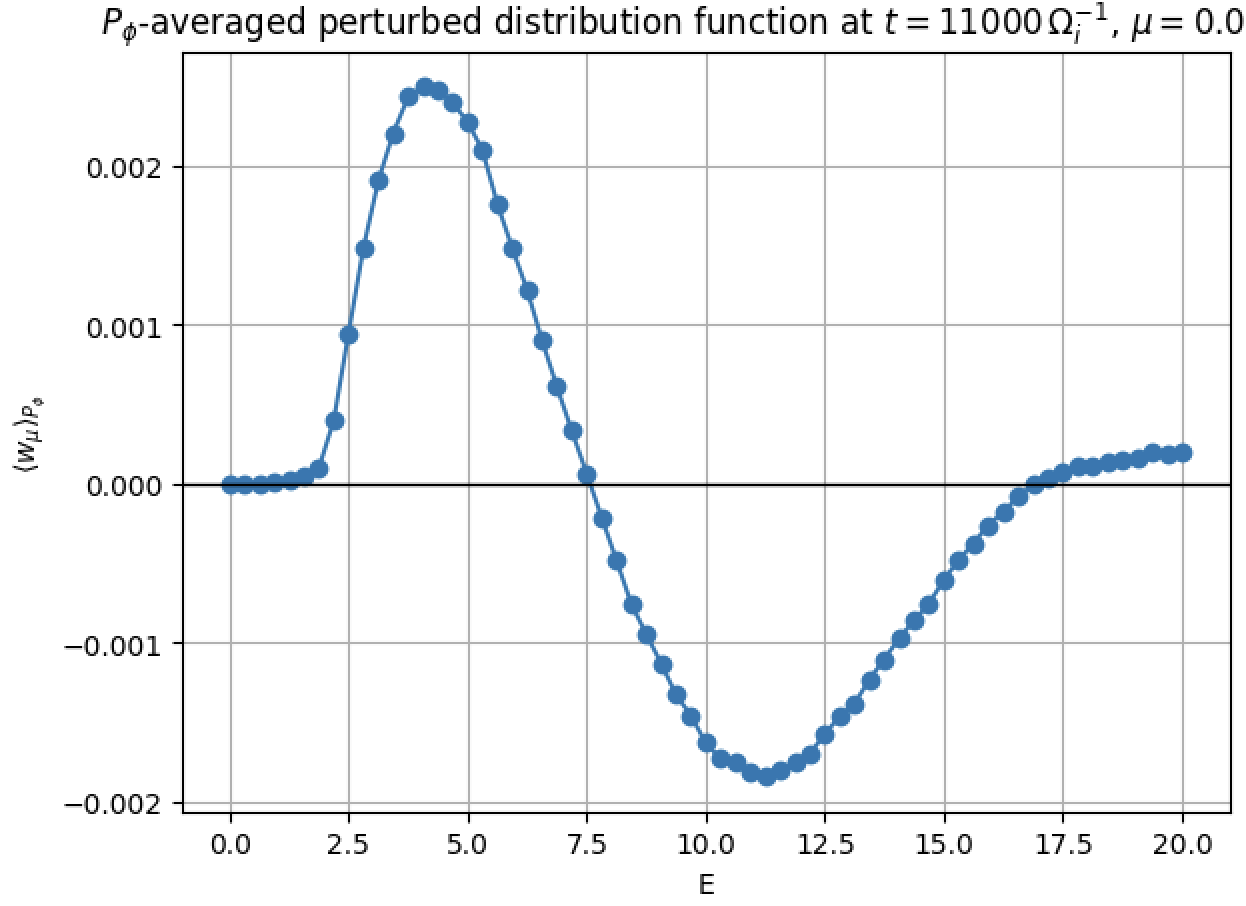}
\caption{$P_\phi$-averaged perturbed distribution function $\langle w \rangle_{P_\phi}(E)$. The resonance energy $E_{\mathrm{res}}$ is determined from the zero-crossing of the averaged profile.}
\label{avg_Pphi}
\end{figure}

To determine the resonance energy more accurately, the perturbed distribution function is further averaged over $P_\phi$. The resulting profile is shown in \autoref{avg_Pphi}.

The $P_\phi$-averaged profile reveals a clear hole–clump-like structure in energy space. The resonance energy $E_{\rm res}$ is identified from the zero crossing separating the positive and negative perturbations, corresponding to the location where the energetic-particle distribution changes sign due to nonlinear wave–particle redistribution.


In the present analysis, the PSZS diagnostic is restricted to the $\mu\simeq0$ slice. This choice corresponds to the small pitch-angle region of phase space, for which the energetic particles are expected to be passing particles. The kinetic energy per unit mass $E$ can therefore be written as $E = v_\parallel^2/2$. The mode frequency can be directly calculated from the resonance energy as $\omega=\frac{v_{\parallel,\mathrm{res}}}{qR} =\frac{\sqrt{2E_{\rm res}}}{qR}$.

\begin{figure}[h]
\centering
\includegraphics[width=0.7\textwidth]{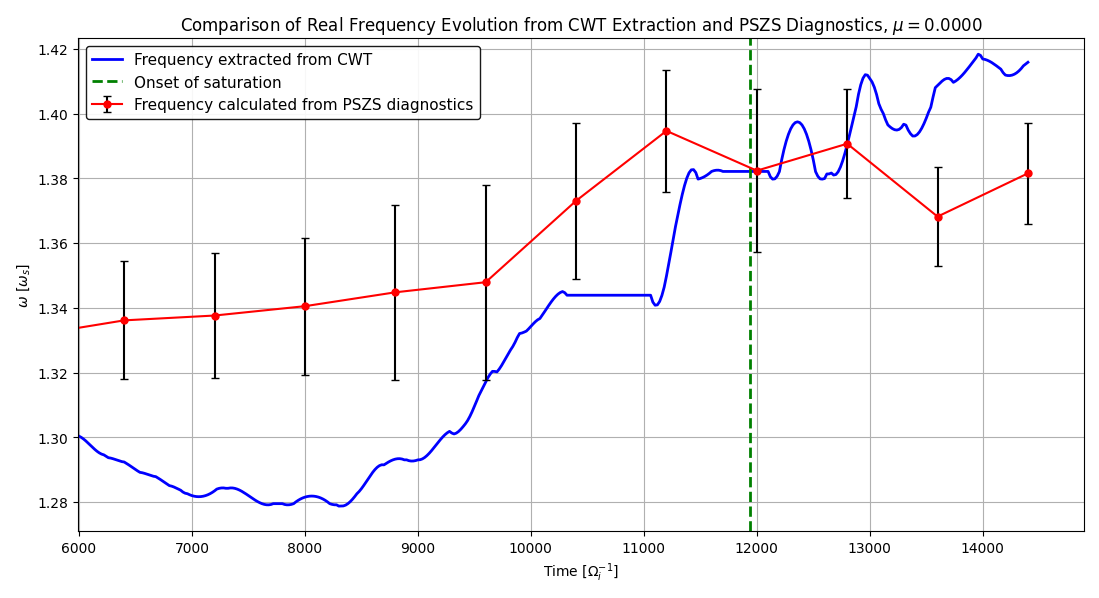}
\caption{Comparison between the mode frequency extracted from the radial electric field using CWT and the frequency reconstructed from the PSZS diagnostics. The vertical dashed green line marks the onset of saturation.}
\label{cwtpszs}
\end{figure}

Applying this procedure to each PSZS snapshot allows the temporal evolution of the mode frequency to be reconstructed and subsequently compared with the frequency obtained from the CWT analysis as shown in \autoref{cwtpszs}.

As shown in \autoref{cwtpszs}, the frequencies extracted from the PSZS diagnostics and from the CWT analysis exhibit a similar chirping behaviour during the nonlinear phase. Focusing on the interval between the onset of nonlinear frequency evolution and the saturation stage (from $t\Omega_i \approx 9000$ to $12000$), both diagnostics capture a comparable increase in the mode frequency. While the absolute frequency values do not coincide exactly, it should be noted that the two diagnostics are based on fundamentally different approaches. The CWT frequency is extracted directly from the temporal evolution of the radial electric field, whereas the frequency derived from the PSZS diagnostics is inferred from the resonance energy identified in phase space through the transit-resonance relation. Despite these differences, the frequency evolution and the corresponding chirping rate are in good agreement. This consistency provides an independent validation of the PSZS-based reconstruction and confirms that the observed frequency chirping is directly linked to the nonlinear evolution of resonant energetic particles in phase space.

%% file: part5.tex
\section{Summary and Discussion}
\label{sec:summary}

Zonal flows play an important role in the regulation of turbulence and transport in tokamak plasmas. A population of energetic particles (EP) can also be found in present tokamak plasmas, and it is predicted to be very important in future reactors. EP-induced geodesic acoustic modes (EGAMs) provide a valuable framework for investigating nonlinear wave--particle interactions and the associated EP phase-space dynamics.

In this work, we have performed a systematic numerical study of EGAMs using the global gyrokinetic particle-in-cell code ORB5. The simulations were carried out in the collisionless electrostatic limit, focusing on a regime where the nonlinear evolution is predominantly governed by wave--particle interactions.
In this framework, the nonlinear evolution of EGAMs is strongly linked to the redistribution of resonant EPs in phase space. As the EP distribution evolves, the drive available for the instability is progressively reduced, leading to saturation. At the same time, the modification of the resonant particle population changes the resonance condition and gives rise to frequency chirping. Understanding the connection between this frequency evolution and the underlying phase-space dynamics constitutes one of the main objectives of the present work.

The simulations show that, after the initial linear phase, EGAMs show a strong nonlinear dynamics, with frequency chirping and a nonlinear saturation. We perform a scan in the EP concentration, to see how the drive has effects on the nonlinear dynamics. With increasing linear growth rate, we observe an increasing saturation level, and the frequency chirping becomes more pronounced. In all cases considered here, the frequency evolution during the early nonlinear phase is characterized by up-chirping. This is consistent with previous experimental measurements investigated in Ref~\cite{Horvath_2016, Lauber_IAEA_2018, Novikau20}. Furthermore, the chirping rate increases systematically with increasing EP drive.

To investigate the origin of the chirping dynamics, we employed the recently developed Phase Space Zonal Structures (PSZS) diagnostics implemented in ORB5. From the perspective of Zonal State dynamics, the observed EGAM saturation and frequency chirping can be interpreted as the coupled evolution of fluctuations and phase-space structures. In this sense, the PSZS diagnostics provide a useful bridge between measurable frequency evolution and the underlying kinetic processes governing energetic-particle transport.

By identifying the resonance energy from the PSZS distribution and reconstructing the corresponding mode frequency, we obtained an independent estimate of the frequency evolution. The reconstructed frequencies exhibit good agreement with those extracted directly from the radial electric field through continuous wavelet transform (CWT) analysis. This correspondence provides direct evidence that the observed frequency chirping is associated with the nonlinear evolution of resonant EP in phase space.

For sufficiently strong EP drive, the simulations reveal substantial redistribution of the EP distribution function extending beyond the immediate resonance layer. In this regime, deviations from the low-drive scaling trends are observed, suggesting that the nonlinear dynamics may involve a broader region of phase space than that described by simplified resonant-particle models. These results indicate that the connection between EGAM chirping and PSZS evolution becomes increasingly complex in the strong-drive regime.

Future work will focus on a more detailed characterization of the underlying phase-space structures, including orbit classification and the identification of the particle populations responsible for the observed chirping behaviour. Such studies will help clarify the role of PSZS in EGAM nonlinear dynamics and EP transport.

\section{Acknowledgments}
The authorsgratefully acknowledges stimulating and fruitful discussions with Alberto Bottino, Alexey Mishchenko, Thomas Hayward-Schneider, Ningfei Chen, and the entire ORB5 team, which provided valuable perspectives and significantly enriched this work. 
This work has been carried out within the framework of the EUROfusion Consortium, partially funded by the European Union via the Euratom Research and Training Programme (Grant Agreement N$^o$ 101052200 — EUROfusion). 
This research was supported by computational resources provided by the Leonardo supercomputer at CINECA.